\documentclass{moriond}

\def\be{\begin{equation}}
\def\ee{\end{equation}}
\def\bea{\begin{eqnarray}}
\def\eea{\end{eqnarray}}

\newcommand{\Photo}{}

\begin{document}

\title{The importance of ln(1/x) resummation: a new QCD analysis of HERA data}

\author{Francesco Giuli}

\address{Particle Physics, Denys Wilkinson Buildingg, Keble Road, University of Oxford, OX1 3RH Oxford, UK}

\maketitle
\abstracts{Fits to the final combined HERA deep-inelastic
scattering cross-section data within the conventional
DGLAP framework of QCD have shown some tension at
low-\textit{x} and low-$Q^{2}$. A resolution of this tension incorporating
$\ln(1/x)$-resummation terms into the HERAPDF fits
is investigated using the {\tt{xFitter}} program. The kinematic
region where this resummation is important is delineated.
Such high-energy resummation not only gives a better description
of the data, particularly of the longitudinal structure
function $F_L$, it also results in a gluon PDF which is
steeply rising at low x for low scales, $Q^{2} \simeq$ 2.7 GeV$^{2}$, contrary
to the fixed-order (FO) NLO and NNLO gluon PDF. This contribution is based on the results presented in Ref. 1~\cite{Abdolmaleki:2018jln}.}
\section{Input data sets}
The input datasets in use are the final combined $e^{\pm}p$ cross-section measurements of H1 and ZEUS~\cite{Abramowicz:2015mha} (both from neutral-current (NC) and charged-current (CC) processes and for $e^+p$ and $e^-p$ scattering) and the HERA combined charm~\cite{Abramowicz:1900rp} from ZEUS and H1. The inclusion of charm data in the fit is useful to determine the optimal charm pole mass. Additionally, since they extend to rather small values of $x$, they may be sensitive to $\ln(1/x)$ resummation effects.
\section{Fit strategy}
\label{sec:setup}
The present QCD analysis uses the {\tt xFitter}
program~\cite{Alekhin:2014irh,h1zeus:2009wt} and is based
on the HERAPDF2.0 setup. The quark distributions at the initial scale $Q_0^2$ were represented by the generic form:
\begin{equation}
  xq_i(x,Q_0) = A_i x^{B_i} (1-x)^{C_i} P_i(x),
\label{eqn:pdf}
\end{equation}
where $P_i(x)=1+\mathcal{O}(x)$ defines a polynomial in powers of $x$. The parametrised quark distributions $q_i$ were chosen to be the valence quark
distributions ($xu_v$, $xd_v$) and the light anti-quark distributions
($x\bar{U}=x\bar{u}$, $x\bar{D}=x\bar{d}+x\bar{s}$). The gluon
distribution was parametrised with the more flexible form:
\begin{equation}\label{eqn:gluonpdf}
xg(x) = A_g x^{B_g} (1-x)^{C_g}P_g(x) - A'_g x^{B'_g} (1-x)^{C'_g}\,.
\end{equation}
The normalisation parameters $A_{u_v}$ and $A_{d_v}$ were fixed using
the quark counting rules and $A_g$ using the momentum sum rule. The
normalisation and slope parameters, $A$ and $B$, of $\bar{u}$ and
$\bar{d}$ were set equal such that $x\bar{u} = x\bar{d}$ at very small
$x$. The strange PDFs $xs$ and $x\bar{s}$ were parametrised as
$xs = x\bar{s}=0.4x\bar{D}$, representing a suppression of strangness
with respect to the light down-type sea quarks, but the input data are
not sensitive to the fraction of strangeness.\\
The $\ln(1/x)$ resummation corrections are available in the {\tt HELL}
code~\cite{Bonvini:2017ogt}, which is a standalone code that implements the resummation corrections to the DGLAP splitting functions $P$ and to the DIS coefficient functions $C$ (both
massless and massive) up to next-to-leading-log accuracy in $\ln(1/x)$
(NLL$x$). The scale at which PDFs are parameterised  have been chosen to be $Q^2_0= 2.56$~GeV$^2$ as compared to 1.9~GeV$^2$ of HERAPDF2.0. The reason is that the numerical computation of $\ln(1/x)$-resummation corrections may become unreliable at low scales due to the large value of the strong coupling
$\alpha_S$.
\section{Results}
The effect of $\ln(1/x)$ resummation on splitting functions and DIS
coefficient functions is more dramatic at NNLO than at NLO~\cite{Bonvini:2017ogt}.
In fact, the full calculation with NNLO+NLL$x$ resummation is closer to the NLO
result than it is to the NNLO result. This is not accidental and is
mostly due to the perturbative instability of the NNLO correction to
the splitting functions generated by small-$x$
logarithms~\cite{Ball:2017otu}. Thus, to better assess the impact of
the $\ln(1/x)$ resummation on the original HERAPDF analysis,
we only focus on NNLO fits.\\
As well as evaluating uncertainties due to the experimental statistical and systematic errors we have perfomed an exploration of model and parametrisation uncertainties as follows. We have varied the charm mass ($m_c=1.41$, $1.51$~GeV), the bottom mass ($m_b=4.25$, $4.75$~GeV), the strong coupling $\alpha_S(m_Z^2)$ ($\Delta \alpha_S=\pm 0.002$), the strangeness fraction ($f_s=0.3$, $f0.5$), the initial scale ($Q^2_0=2.88$~GeV$^2$), and the $Q^2$ cut on the data ($Q^2_{\rm min}= 2.7$~GeV$^2$, $5$~GeV$^2$).\\
Furthermore, parametrisation uncertainties have been explored by adding extra terms to the polynomials $P_i(x)$ of Eq.~\ref{eqn:pdf}. The only noticeable difference comes from  the addition of a linear term to the polynomial $P_{u_V}(x)$ of the valence up quark PDF.
The largest contribution to the uncertainty on the gluon distribution arises from the variation of the $Q^2_{\rm min}$ cut to $5$~GeV$^2$. Interestingly, this uncertainty is reduced for the fit with $\ln(1/x)$ resummation, due to reduced tensions with the data. 
\begin{figure*}[t]
  \begin{centering}
    \includegraphics[width=0.32\textwidth]{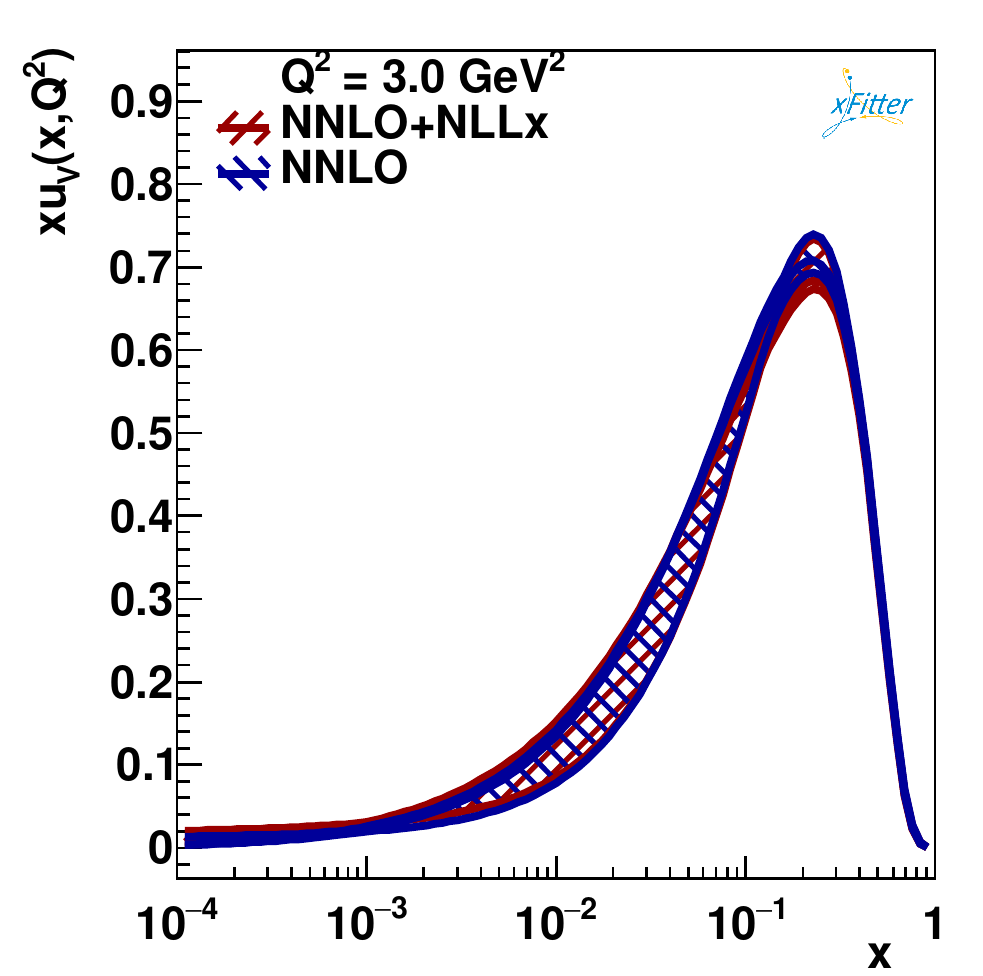}
    \includegraphics[width=0.32\textwidth]{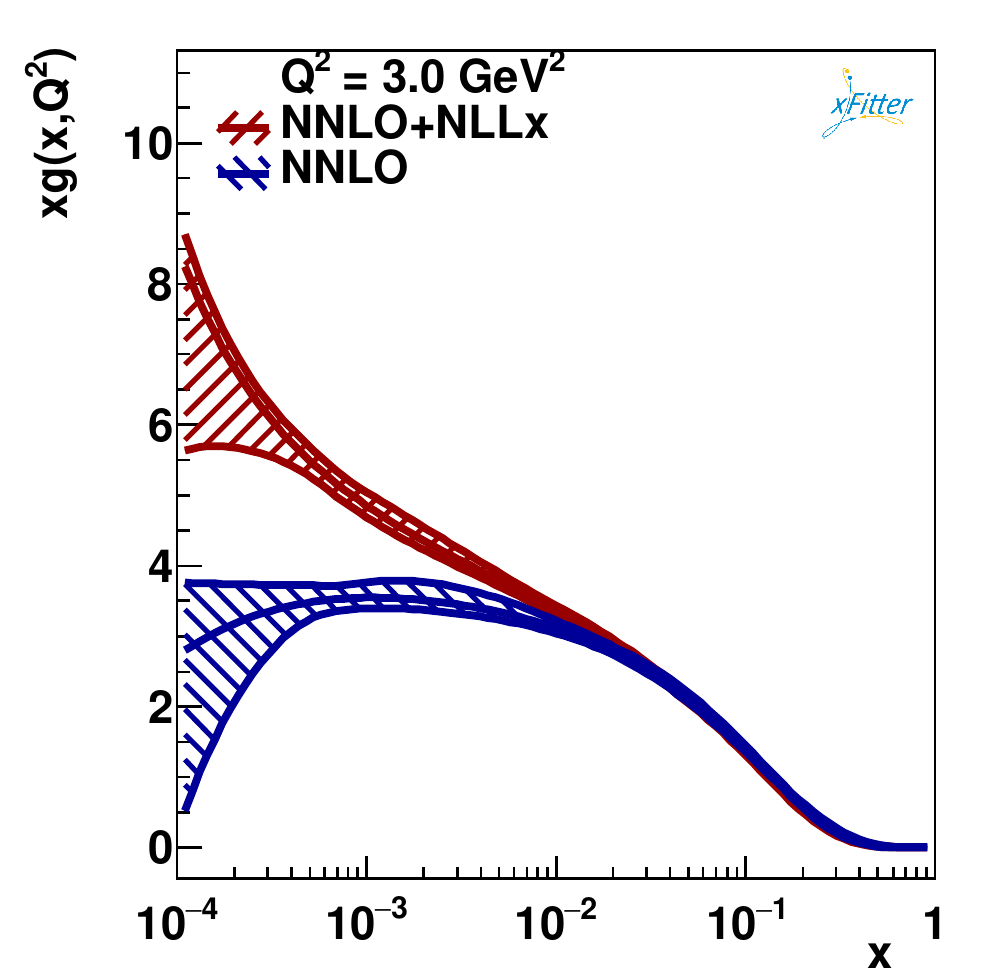}
    \includegraphics[width=0.32\textwidth]{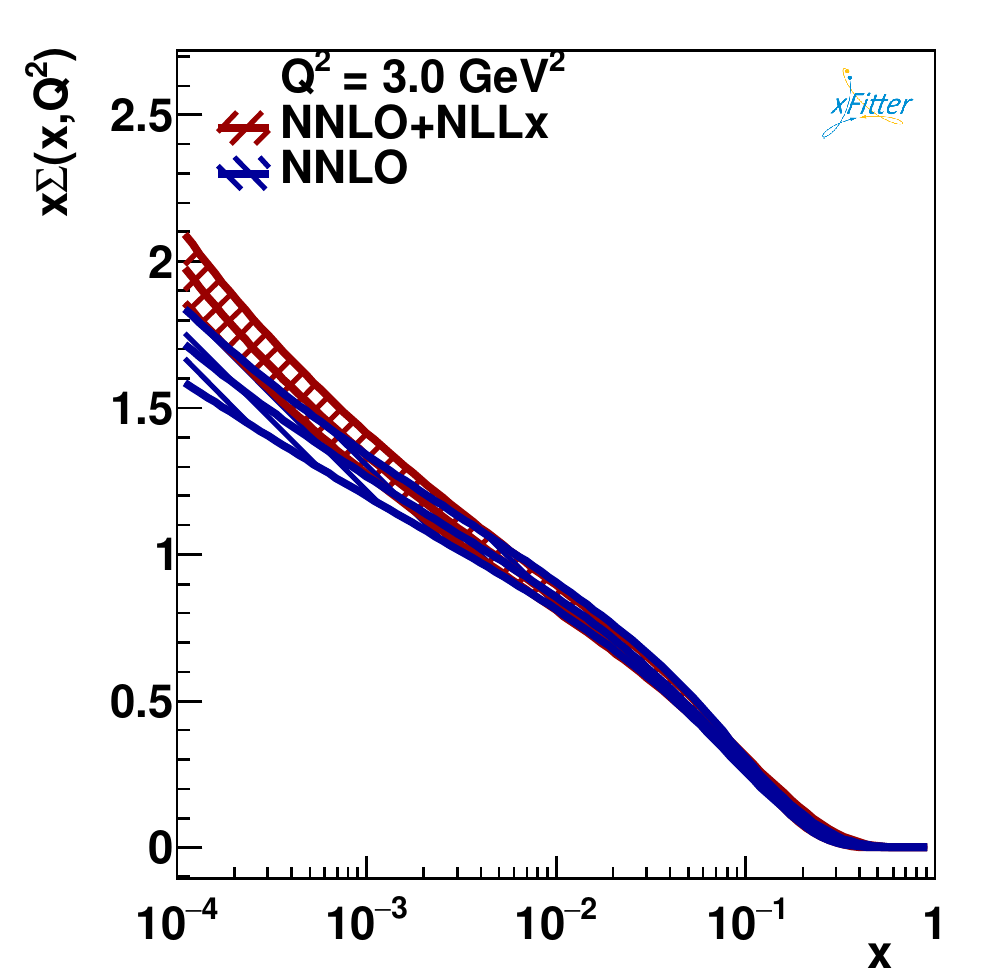}
    \caption{The up valence PDF $xu_v$, the gluon PDF $xg$ and the
      total singlet PDF $x\Sigma$ for the final fits with
      (NNLO+NLL$x$) and without (NNLO) $\ln(1/x)$
      resummation.}
    \label{fig:finalpdf}
  \end{centering}
\end{figure*}
Fig.~\ref{fig:finalpdf} shows a comparison of PDFs
with and without $\ln(1/x)$ resummation at $Q^2 = 3$~GeV$^2$.
This figure displays also the full uncertainty bands. When resummation is included, both the gluon and the total singlet PDFs rise towards low $x$, in contrast to the behaviour of the gluon when resummation is not included.\\
\begin{table*}\small
\begin{center}
\begin{tabular}{cccccc}
  \hline
  &NNLO fit &  NNLO+NLL$x$ fit &   & &   \\
  &  with new settings      &  with new settings  &  &      & \\
  \hline
  Total $\chi^2/\rm{d.o.f}$ & 1446/1178 & 1373/1178& & & \\
  correlated $\tilde\chi^2/\rm{n.d.p}$ inclusive & 102 & 77 & & & \\
  correlated $\tilde\chi^2/\rm{n.d.p}$ charm& 15 & 11 & & & \\
  log $\tilde\chi^2/\rm{n.d.p}$ inclusive& 20 & $-$3 & & &\\
  log $\tilde\chi^2/\rm{n.d.p}$ charm & $-$2 & $-$1 & & &\\
  \hline
\end{tabular}
\caption{Total $\chi^2$ per d.o.f. for the PDF fits to HERA inclusive
  and charm data with the new settings. Also shown are the
  contributions to the $\tilde\chi^2/\rm{n.d.p}$ from the correlated shifts and the log
  terms.}
\label{tab:fitresults2}
\end{center}
\end{table*}
The $\chi^2$ values for the fits are summarised in
Tab.~\ref{tab:fitresults2}. There is a decrease of $73$ units in
$\chi^2$ when the $\ln(1/x)$ resummation is used. Most of this
difference is coming from the highly accurate NC $E_p = 920$~GeV data
which probe the low-$x$ and low-$Q^2$ region and are thus most
sensitive to $\ln(1/x)$ resummation (413/377 to be compared to 446/377). 
As expected, a decrease in $\chi^2$ has been also observed in the NC
$E_p=820$~GeV (65/70 for the NNLO+NLLx fit vs. 70/70 for the FO NNLO fit) and in the charm data (49/47 vs. 48/47), which are also expected to have some sensitivity. Other data sets entering the fit probe higher $x$ and $Q^2$ and their $\chi^2$ are not significantly changed. 
\begin{figure}[t]
  \begin{centering}
    \includegraphics[width=0.32\textwidth]{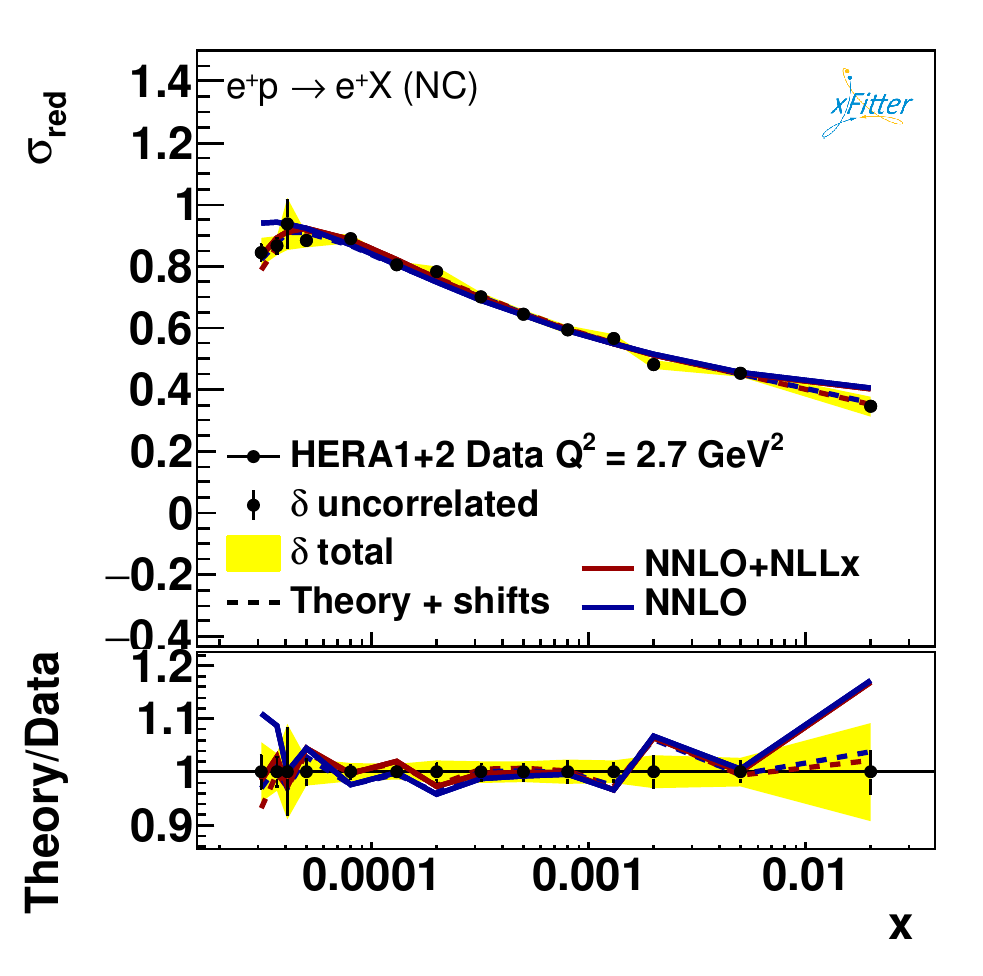}
    \includegraphics[width=0.32\textwidth]{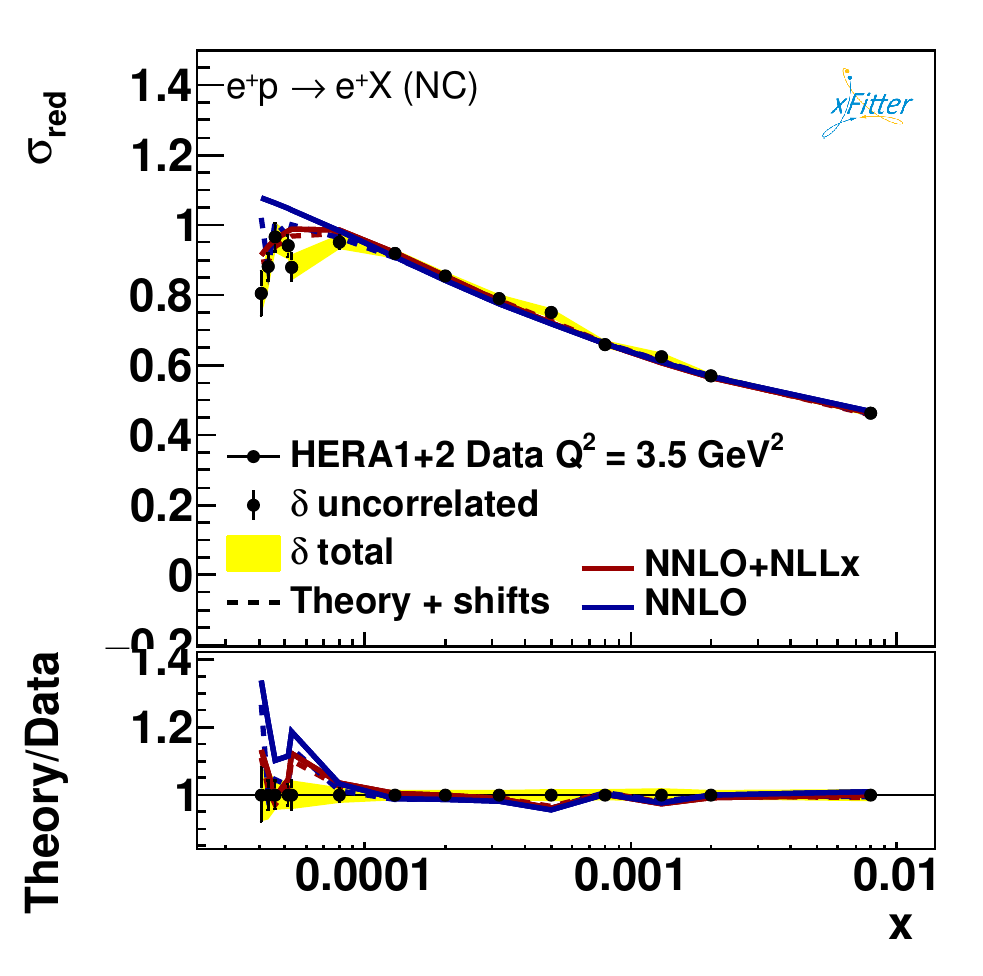}
    \includegraphics[width=0.32\textwidth]{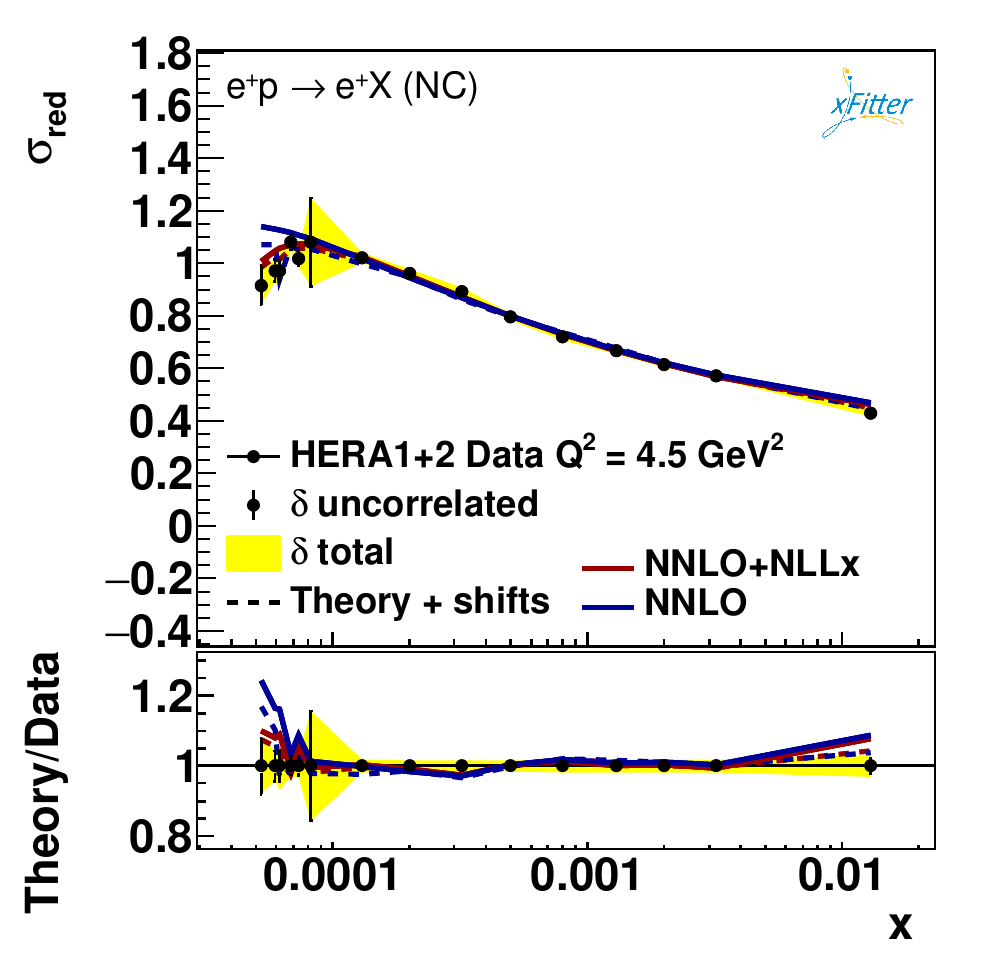}
    \caption{The HERA NC $E_p= 920$~GeV data compared to the fits with
      and without $\ln(1/x)$ resummation for the $Q^2 = 3.5$, $Q^2 = 3.5$ and $4.5$~GeV$^2$ bins.\label{fig:data920}}
  \end{centering}
\end{figure} 
In Fig.~\ref{fig:data920} the fit results are compared to the NC
$E_p=920$~GeV inclusive reduced cross-section data in the lowest $Q^2$
bins included in the fits. It is evident that for the fit including
$\ln(1/x)$-resummation effects, not only the initial description of
the data is better, but also the correlated shifts are smaller and this
is one of the reasons why the $\chi^2$ of the fit is significantly
smaller. In particular, it is evident that the low-$x$ turn-over of
the measurements is better reproduced by the fit that includes
$\ln(1/x)$ resummation, which in turn explains the big reduction of the $\chi^2$.
This is a direct consequence of the steeper
gluon at low $x$ (see Fig.~\ref{fig:finalpdf}) which makes $F_L$
larger at low $x$ causing a more pronounced turn-over of the reduced
cross section, defined as follows: 
\begin{equation}\label{eq:redxsec}
  \sigma_{\rm red} =F_2 - \frac{y^2}{Y_{+}} F_L\,,
\end{equation}
where $F_2$ and $F_L$ are the structure functions related to the parton distributions~\cite{Gao:2017yyd}, $Y_{+}= 1 + (1-y)^2$ and $y = Q^{2}/(sx)$.
\begin{figure}[t]
  \begin{centering}
    \includegraphics[width=0.49\textwidth]{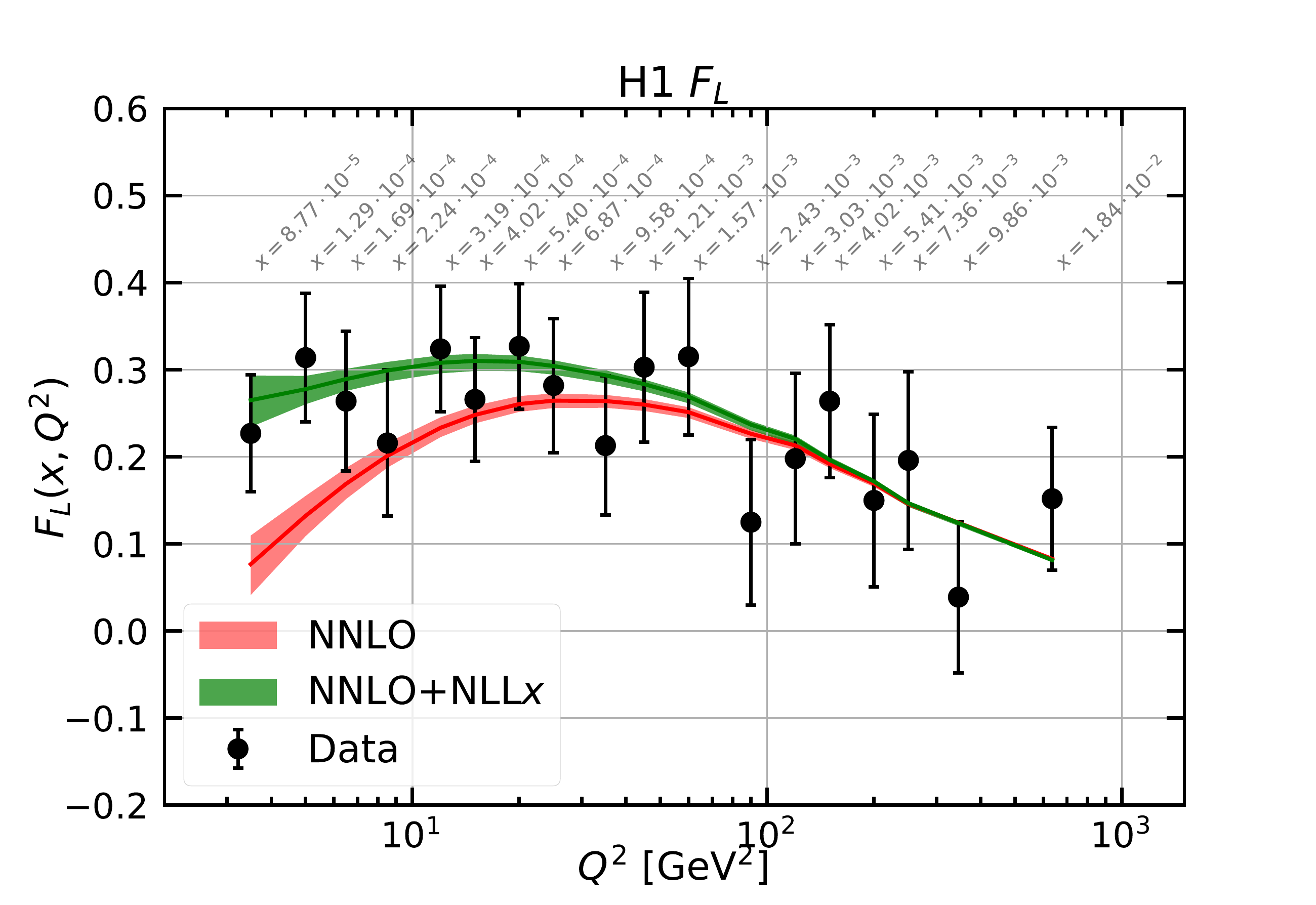}
    \includegraphics[width=0.49\textwidth]{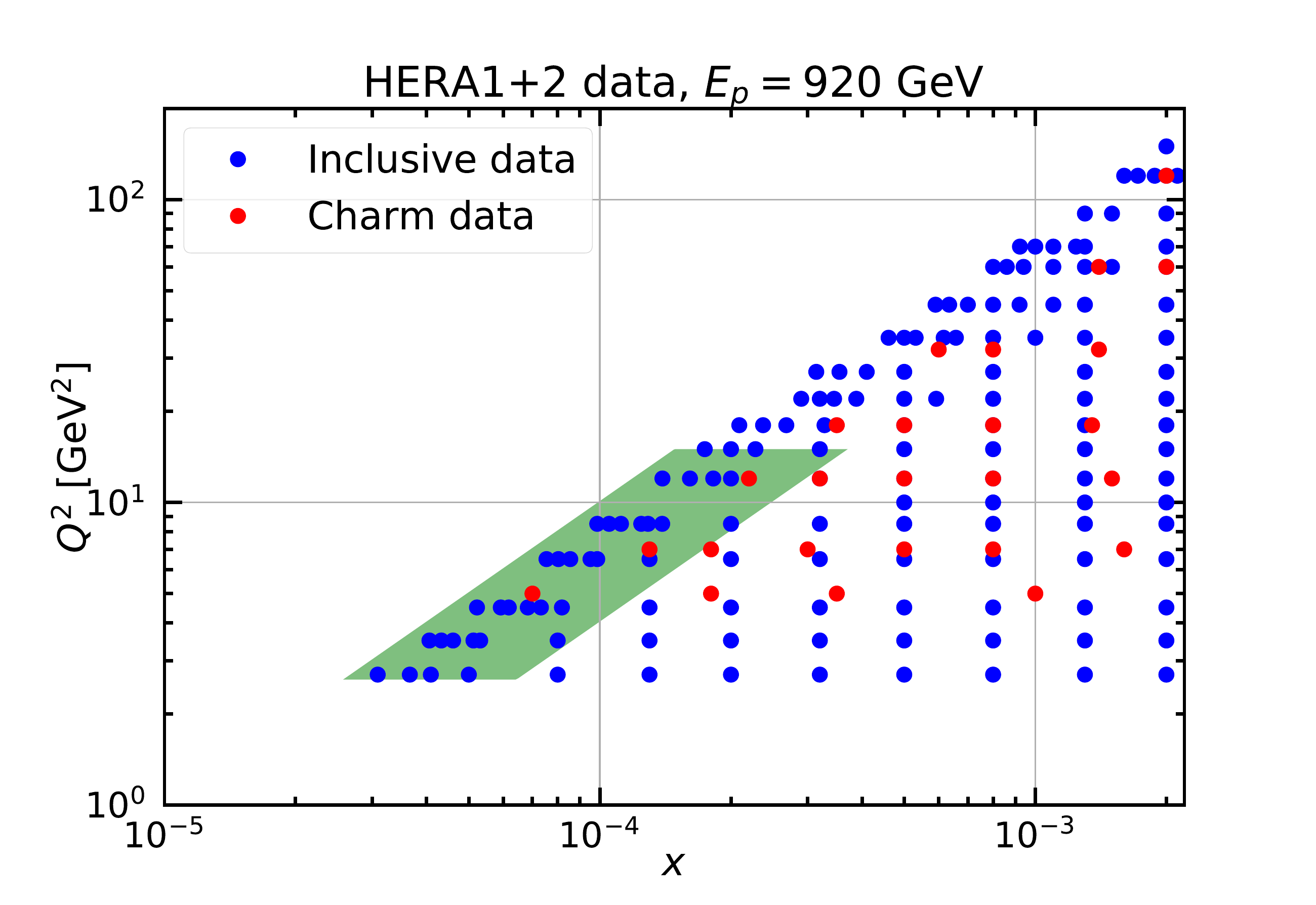}
    \caption{Left: The H1 measurement of $F_L$ compared to the
      predictions with and without $\ln(1/x)$
      resummation; Right: Scatter plot of the low-$x$ and low-$Q^2$ kinematic
      region covered by the HERA1+2 inclusive data and charm data at
      $E_p= 920$~GeV. The green shaded area indicates the region in
      which $\ln(1/x)$ resummation has a significant effect.}
      \label{fig:f2fl}
  \end{centering}
\end{figure}
This point is also illustrated in Fig.~\ref{fig:f2fl} where the H1 $F_L$
measurement is compared to the theoretical predictions
of $F_L$ with and without $\ln(1/x)$. It is clearly visible that the description of this data set is improved in the former case thanks to the fact that $\ln(1/x)$-resummed
predictions for $F_L$ are larger at low $x$.\\
The results presented so far indicate that the improvement of the
description of the HERA data when including $\ln(1/x)$ resummation is
driven by the low-$x$ and low-$Q^2$ data. We can also delineate
the kinematic region responsible for the improvement more
precisely. To do so, we have performed $\chi^2$ scans in
$Q^2_{\rm min}$ with no cut in $x$, and in $x_{\rm min}$ (where
$x_{\rm min}$ is the minumum value of Bjorken $x$ allowed in the fit)
fixing $Q^2_{\rm min} = 2.7$~GeV$^2$. Furthermore, an additional $\chi^2$ scan in $y_{\rm max}$ has been done, excluding from the fit data with $y > y_{\rm max}$. The $\chi^2$ scans as a function of $Q_{\rm min}$, $x_{\rm min}$ and $y_{\rm max}$ allow us to delineate the region of the $(x,Q^2)$-plane in which $\ln(1/x)$ resummation is important.\footnote{The actual plane over which the constraint acts is the $(x,Q^2/s)$-plane. However, for simplicity in the following we will only consider the $E_p=920$~GeV inclusive and the charm datasets that were both taken at $\sqrt{s}=318$~GeV.} Fig.~\ref{fig:f2fl} displays a zoom of the low-$x$ and low-$Q^2$ kinematic region covered by the HERA1+2 inclusive and charm data at $E_p= 920$~GeV. The green shaded area indicates the region such that $x<5\cdot 10^{-4}$, 2.7~GeV$^2 < Q^2 < 15$~GeV$^2$, and $0.4 < y < 1$ (assuming $\sqrt{s} = 318$~GeV) determined by combining the results of the scans discussed above.\footnote{In fact, given the range in $y$, the constraint on $x$ has no effect on the shaded area.} This provides an estimate of the region where $\ln(1/x)$ resummation provides a significantly better description of the HERA data as compared to FO predictions.
\section{Conclusion}
In conclusion, $\ln(1/x)$ resummation provides a substantial
improvement in the description of the precise HERA1+2 combined data and it overcomes a major disadvantage of the FO analyses, namely a decreasing gluon PDF at low $x$ and $Q^2$. It represents an alternative to the addition of higher-twist terms~\cite{Abt:2016vjh,Harland-Lang:2016yfn,Motyka:2017xgk} and does not suffer from the pathological features of some of these analyses~\cite{Abt:2016vjh}.

\end{document}